\documentclass[]{aa}

\usepackage[utf8]{inputenc}
\usepackage{float}
\usepackage{natbib}
\usepackage{graphicx}
\usepackage{subcaption}
\usepackage{numprint}
\usepackage{color}
\usepackage{txfonts}
\usepackage[colorlinks = true, allcolors = blue]{hyperref}
\usepackage{tabularx}
\usepackage{amsmath}
\usepackage{amssymb}
\usepackage{siunitx}
\usepackage{mathtools}
\usepackage[normalem]{ulem}
\usepackage{tikz}
\usetikzlibrary{calc,shapes,arrows}
\usepackage{multirow}
\usepackage{threeparttable}
\begin{document}

\title{On the Milky Way spiral arms from open clusters in \textit{Gaia} EDR3}

\author{
    A. Castro-Ginard          \inst{\ref{inst:UB}}\relax
\and  P.J. McMillan         \inst{\ref{inst:lund}}\relax
\and  X. Luri         \inst{\ref{inst:UB}}\relax
\and    C. Jordi        \inst{\ref{inst:UB}}\relax
\and   M. Romero-G\'omez        \inst{\ref{inst:UB}}\relax
\and   T. Cantat-Gaudin       \inst{\ref{inst:UB}}\relax
\and   L. Casamiquela       \inst{\ref{inst:bordeaux}}\relax
\and   Y. Tarricq       \inst{\ref{inst:bordeaux}}\relax
\and   C. Soubiran        \inst{\ref{inst:bordeaux}}\relax
\and   F. Anders        \inst{\ref{inst:UB}}\relax
}

\institute{Dept. F\'isica Qu\`antica i Astrof\'isica, Institut de Ci\`encies del Cosmos (ICCUB), Universitat de Barcelona (IEEC-UB), Mart\'i i Franqu\`es 1, E08028 Barcelona, Spain\\
    \email{acastro@fqa.ub.edu}\relax \label{inst:UB}
\and
Lund Observatory, Department of Astronomy and Theoretical Physics, Lund University, Box 43, SE-22100, Lund, Sweden \relax \label{inst:lund}
\and
Laboratoire d'Astrophysique de Bordeaux, Univ. Bordeaux, CNRS, B18N, all\'{e}e Geoffroy Saint-Hilaire, 33615 Pessac, France \relax \label{inst:bordeaux}
}

\date{Received date /
Accepted date}

\abstract{
    The physical processes driving the formation of Galactic spiral arms are still under debate. Studies using open clusters favour the description of the Milky Way spiral arms as long-lived structures following the classical density wave theory. Current studies comparing the {\em Gaia} DR2 field stars kinematic information of the Solar neighbourhood to simulations, find a better agreement with short-lived arms with a transient behaviour.
}{
  	Our aim is to provide an observational, data-driven view of the Milky Way spiral structure and its dynamics using open clusters as the main tracers, and to contrast it with simulation-based approaches. We use the most complete catalogue of Milky Way open clusters, with astrometric {\em Gaia} EDR3 updated parameters, estimated astrophysical information and radial velocities, to re-visit the nature of the spiral pattern of the Galaxy.
}{
    We use a Gaussian mixture model to detect overdensities of open clusters younger than $30$ Myr that correspond to the Perseus, Local, Sagittarius and Scutum spiral arms, respectively. We use the birthplaces of the open cluster population younger than $80$ Myr to trace the evolution of the different spiral arms and compute their pattern speed. We analyse the age-distribution of the open clusters across the spiral arms to explore the differences in the rotational velocity of stars and spiral arms.
}{
    We are able to increase the range in Galactic azimuth where present-day spiral arms are described, better estimating its parameters by adding $264$ young open clusters to the $84$ high-mass star forming regions used so far, thus increasing by a $314\%$ the number of tracers. We use the evolution of the open clusters from their birth positions to find that spiral arms nearly co-rotate with field stars at any given radius, discarding a common spiral pattern speed for the spiral arms explored.
}{
	The derivation of different spiral pattern speeds for the different spiral arms disfavours classical density waves as the main drivers for the formation of the Milky Way spiral structure, and is in better agreement with simulation-based approaches that tend to favour transient spirals. The increase in the number of known open clusters, as well as in their derived properties allows us to use them as effective spiral structure tracers, and homogenise the view from open clusters and field stars on the nature of the Galactic spiral arms.
}
\keywords{Galaxy: disc — open clusters and associations: general — astrometry — Methods: data analysis} 
\maketitle


\section{Introduction}
\label{sec:intro}

The location of our Solar system within the Milky Way disc makes it challenging to obtain a detailed picture of its structure. This is particularly true for the spiral structure. The number and location of the spiral arms still remains unclear, as well as their nature. \citet{1964ApJ...140..646L} proposed a theoretical mechanism for the formation of spiral arms, widely known as the density wave theory, where the spiral arms rotate like a rigid solid at a constant angular velocity (\textit{i.e.} pattern speed) in spite of the differential rotation of the stars and interstellar medium, causing the spiral arms to be long-lived \citep[see ][for a review of the classic theory]{2016ARA&A..54..667S}. Alternatively, \citet{1964ApJ...139.1217T} proposed that spiral arms could be reforming short-lived structures composed by individual arms, each of them behaving as a wave at a constant pattern speed, which overlap causing transient spiral arms with no global spiral pattern speed \citep{2011MNRAS.417..762Q,2014ApJ...785..137S}. This latter short-lived arms can be also explained with material arms that co-rotate with disc stars \citep{2011ApJ...735....1W,2012MNRAS.421.1529G}, causing the spiral pattern to grow from local gravitational instabilities and then to disappear, with continuous instabilities regenerating the pattern again.

Since the first attempts to explain the nature of the arms almost $60$ years ago, no clear conclusion has been reached. \citet{2014PASA...31...35D} proposed types of observational evidence to shed light on the nature of the spiral structure, based on the different rotation velocities for the spiral pattern and disc stars. These strategies include the direct derivation of the spiral pattern speed for each arm, which will help in favouring either a density wave theory, where the arms share a global constant spiral pattern speed, or a transient behaviour which shows a spiral pattern speed decreasing with Galactocentric radius \citep{2018MNRAS.478.3590S}. The distribution of ages of the stellar clusters across any spiral arm also indicates a difference in the velocities of both structures, the distribution of stars and the arm. Given the improvements on the open cluster catalogue made in light of {\em Gaia} second data release \citep[{\em Gaia} DR2, ][]{2018A&A...616A...1G}, and using them as main tracers of Galactic spiral structure, both observational evidences can be pursued for the first time providing a new view of the nature of the Milky Way spiral arms.

Open clusters (OCs) are excellent tracers of the spatial structure of the young stellar population in the Galactic disc. They are groups of stars, gravitationally bound, which were born from the same molecular cloud and, therefore, have very similar positions, velocities, ages and chemical composition \citep{2003ARA&A..41...57L}. For an OC, the estimation of its properties such as the parallax, proper motion, radial velocity, age or extinction is more reliable than for individual field stars because it relies on a (large) number of members, whose parameters are averaged. 

Using OCs as spiral arms main tracers, \citet{2007NewA...12..410N} found that spiral pattern speeds for the Perseus, Local and Sagittarius spiral arms decrease with Galactocentric radius, finding evidence for multiple spiral sets. Also using young OCs as spiral arms main tracers and {\em Gaia} DR2, \citet{2019MNRAS.486.5726D} obtained a common pattern speed for the Perseus, Local and Sagittarius spiral arms of $28.2 \pm 2.1$ km s$^{-1}$ kpc$^{-1}$, supporting the idea of the density wave nature of the spiral structure. This results in a co-rotation radius, \textit{i.e.} Galactocentric radius at which the spiral pattern speed coincides with the velocity from the Galactic rotation curve, of $R_c = 8.51 \pm 0.64$ kpc. \citet{2015MNRAS.449.2336J} used a sample of giant stars from OCs observed by APOGEE DR10 \citep{2014A&A...564A.115A} to find a pattern speed of $23.0 \pm 0.5$ km s$^{-1}$ kpc$^{-1}$, with a corresponding co-rotation radius of $R_c = 8.74$ kpc, compatible to the previous result within uncertainties. However, even though $R_c$ is a fundamental parameter in the density wave scenario, there is not a consensus on its value yet. Different studies, using different tracers, place it from $6.7$ kpc to beyond the Perseus arm, located at $\sim 10$ kpc \citep{2001ApJ...556..181D,2015A&A...577A.142M,2018ApJ...863L..37M}.

From a complementary point of view, by comparing the kinematic substructure of field stars in the Solar neighbourhood to simulated data, \citet{2018MNRAS.481.3794H} showed that a simulated Galaxy with transient spiral arms reproduces the arches and ridges seen in the velocity distribution of {\em Gaia} DR2 \citep{2018A&A...616A..11G,2018Natur.561..360A,2018A&A...619A..72R}. In this transient scenario $R_c$ is not an important parameter since the spiral arms would co-rotate with stars at their Galactocentric radius, causing short-lived arms. A number of authors looking at the field population, some using simulation-based approaches, tended to favour a transient nature for the spirals \citep{2018MNRAS.480.3132Q,2020MNRAS.497..818H,2020arXiv200710990K}.

Since the publication of the {\em Gaia} DR2, hundreds of new OCs have been detected \citep{acastro1,acastro2,2020A&A...635A..45C,2019ApJS..245...32L,upk_clusters} and have been added to the OCs known before (and confirmed by) {\em Gaia} DR2 \citep{tristan_catalogue}. For this compendium of OCs, information about age, distance and line-of-sight extinction was computed \citep{2020A&A...640A...1C} and radial velocities were compiled from different ground-based spectroscopic surveys \citep{2020arXiv201204017T}. Altogether results in a robust OCs catalogue that offers the chance to trace the spiral structure of the Galactic disc (in the Solar neighbourhood) and its evolution over the past few hundred Myr.

Our aim for this paper is to use this recent and homogeneous OC sample with information on astrometric mean parameters, radial velocities and astrophysical parameters available to discriminate as far as possible among different theories for the nature of the spiral structure of the Milky Way, supporting either classical density waves or transient spiral arms.

The paper is organised as follows. In Sect.~\ref{sec:data} we describe the OC sample that we use throughout the analysis. In Sect.~\ref{sec:present_age} we study the spatial distribution of the reported OCs, particularly the youngest ones, and derive the present-day spiral arms structure. In Sect.~\ref{sec:spiral_patspd} we use the astrophysical information of OCs (\textit{i.e.} phase-space coordinates and ages) to test the density wave nature of the spiral arms, by computing the spiral pattern speed for the Perseus, Local, Sagittarius and Scutum spiral arm segments. In Sect.~\ref{subsec:age_distr} we explore the imprints left in the age-distribution of the open clusters across the spiral arms, by the differences in the rotational velocity of the stars and the arms. The discussion on the results obtained is done in Sect.~\ref{sec:discussion}, and the conclusions are found in Sect.~\ref{sec:conclusions}.

\section{The open cluster sample}
\label{sec:data}

The data used throughout the paper are those from the OCs identified in the \textit{Gaia} DR2 data \citep{2018A&A...616A...1G}, with their mean astrometric values updated with {\em Gaia} EDR3 measurements \citep{2020arXiv201201533G}. The use of OCs allows us to have better constrained parameters than using field stars. The parameters needed for our methodology are the mean astrometric parameters, \textit{i.e.} $(l,b,\mu_{\alpha^*},\mu_\delta)$; the astrophysical parameters derived from {\em Gaia} astrometry and photometry, \textit{i.e.} OC age, distance and line-of-sight extinction \citep{2020A&A...640A...1C}; and radial velocity measurements for each OC \citep{2020arXiv201204017T}.

\subsection{\textit{Gaia} EDR3 astrometry}
\label{subsec:open_clusters_astrometry}

The sample of OCs used in this work includes those known previous to (and confirmed by) \textit{Gaia} DR2 \citep{tristan_catalogue}. Additionally, we have included the large number of clusters which have been found in {\it Gaia} DR2 data \citep[\textit{e.g.}][]{acastro1,acastro2,2020A&A...635A..45C,upk_clusters,2019ApJS..245...32L}. We updated the OC mean astrometric parameters with the {\em Gaia} EDR3 astrometric information, which has a greater precision in its measurements given the longer time baseline for the observations. In total there are $2017$ OCs in these catalogues. For those clusters we use the sky coordinates and proper motions for the centre of the OC $(l,b,\mu_{\alpha^*},\mu_\delta)$, which are computed from its member stars. The uncertainties in the position $(l,b)$ can be neglected. The uncertainties in the mean proper motions are below $0.2$ mas yr$^{-1}$, generally around $0.1$ mas yr$^{-1}$.

\subsection{Age, distance and line-of-sight extinction}
\label{subsec:ages}

\citet{2020A&A...640A...1C} published a catalogue of ages, distances and line-of-sight extinctions for the OCs known to date. They used an artificial neural network to infer these parameters for each OC from its colour-magnitude diagram in the \textit{Gaia} passbands $(G,G_{BP},G_{RP})$ and parallax information $(\varpi)$. The authors were able to compute these astrophysical parameters for $1878$ of the $2017$ OCs reported. For the $139$ others, the OC had too few members or its CMD was too red, and no reliable estimation could be obtained using {\it Gaia} data alone.

The uncertainties in those quantities depend on the number of cluster members \citep[see Sect.~3.4 of][for details]{2020A&A...640A...1C}. We took the one sigma uncertainty for the age as $\sigma_{\log{t}} \in [0.15,0.25]$ dex, which are the values for the young OCs uncertainties recommended by \citet{2020A&A...640A...1C}. The uncertainty on the distance modulus is within $0.1$ to $0.2$ mag, corresponding to a $5\%$-$10\%$ distance uncertainty. \citet{2020A&A...640A...1C} reported that they found no systematic effects on the determination of the parameters with respect to the literature. 

\subsection{Radial velocities}
\label{subsec:rad_vel}

Radial velocities used in this work are those compiled by \citet{2020arXiv201204017T}. The authors crossmatched the OC members with several radial velocity catalogues. In addition to {\em Gaia}-RVS \citep{2019A&A...622A.205K}, they used data from ground-based large spectroscopic surveys: the latest public version of the {\em Gaia}-ESO survey \citep{ges}, APOGEE DR16 \citep{apo}, RAVE DR6 \citep{2020AJ....160...82S}, GALAH DR2 \citep{galah,zwi18}. They also included data from other radial velocity catalogues: \citet{nor04}, \citet{mer08,mer09}, \citet{wor12}, the OCCASO survey \citep{casa2016} and \citet{sou18b}.

This catalogue consists of radial velocity estimates for $1382$ clusters, with $1321$ of them having astrophysical parameters estimated by \citet{2020A&A...640A...1C}. The median radial velocity uncertainty of the full catalogue is $1.13$ km s$^{-1}$. We refer to the original paper for the details in the computation of the cluster radial velocities.


\subsection{Final OC sample}
\label{subsec:final_sample}

\begin{figure}[!htb]
\includegraphics[width = 1.\columnwidth]{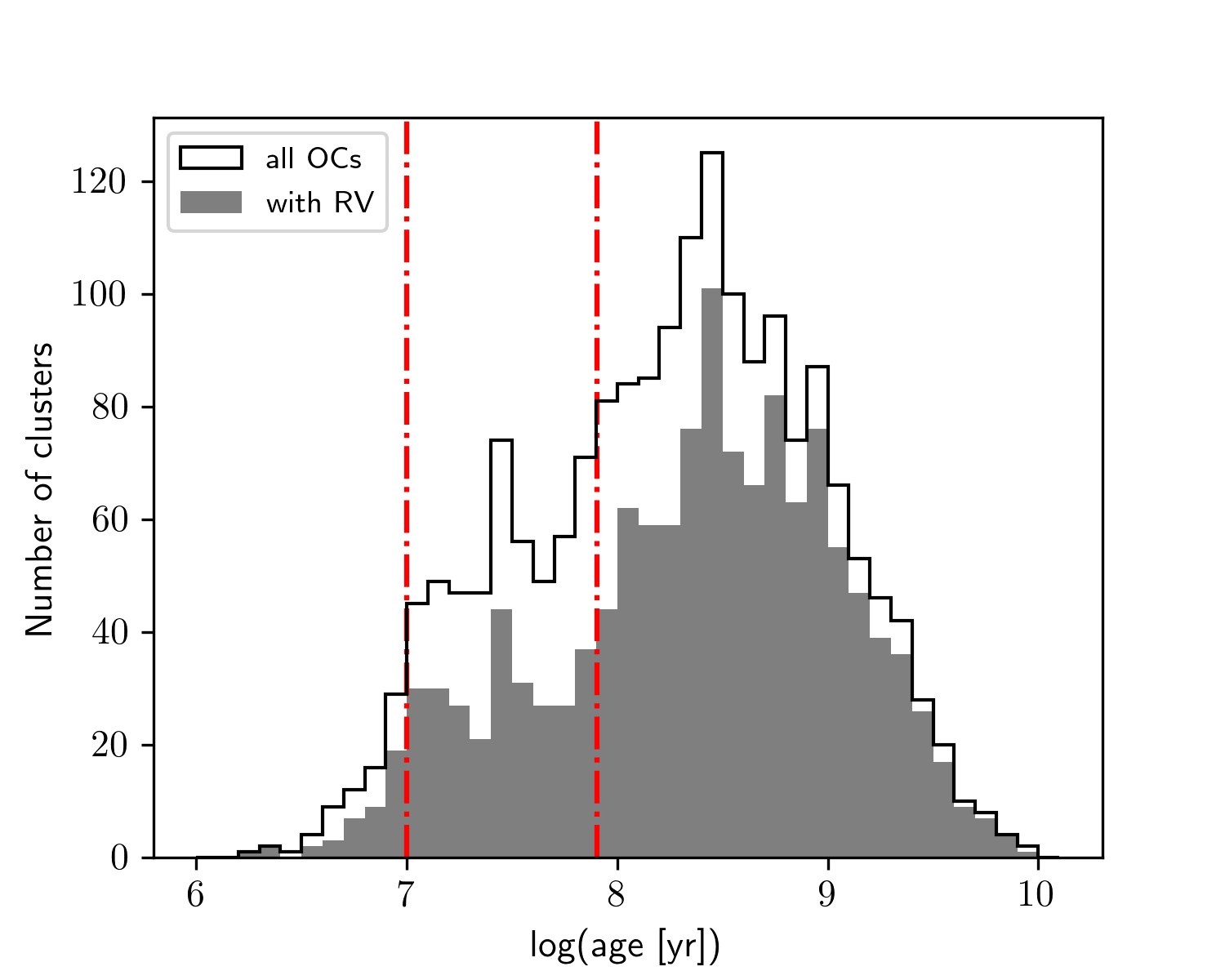}
\caption{Histogram of OC ages. Solid line shows OCs with age estimation available whilst solid bars show the subset of OCs with radial velocity measurements. Red dash-dotted vertical lines correspond to $10$ and $80$ Myr.}
\label{fig:OC_sample}
\end{figure}

Figure~\ref{fig:OC_sample} shows the distribution of ages of the two subsamples of OCs, the one with age, distance and line-of-sight extinction determination is represented as a solid black line; whilst the grey bars represent the clusters with radial velocity measurements. The red dash-dotted lines correspond to ages equal to $10$ and $80$ Myr, and they show the subset of OCs that we use to compute the spiral pattern speed in Sect.~\ref{sec:spiral_patspd}. 

\section{Present-day OC spatial distribution}
\label{sec:present_age}

\begin{figure}[!htb]
\includegraphics[width = 1.\columnwidth]{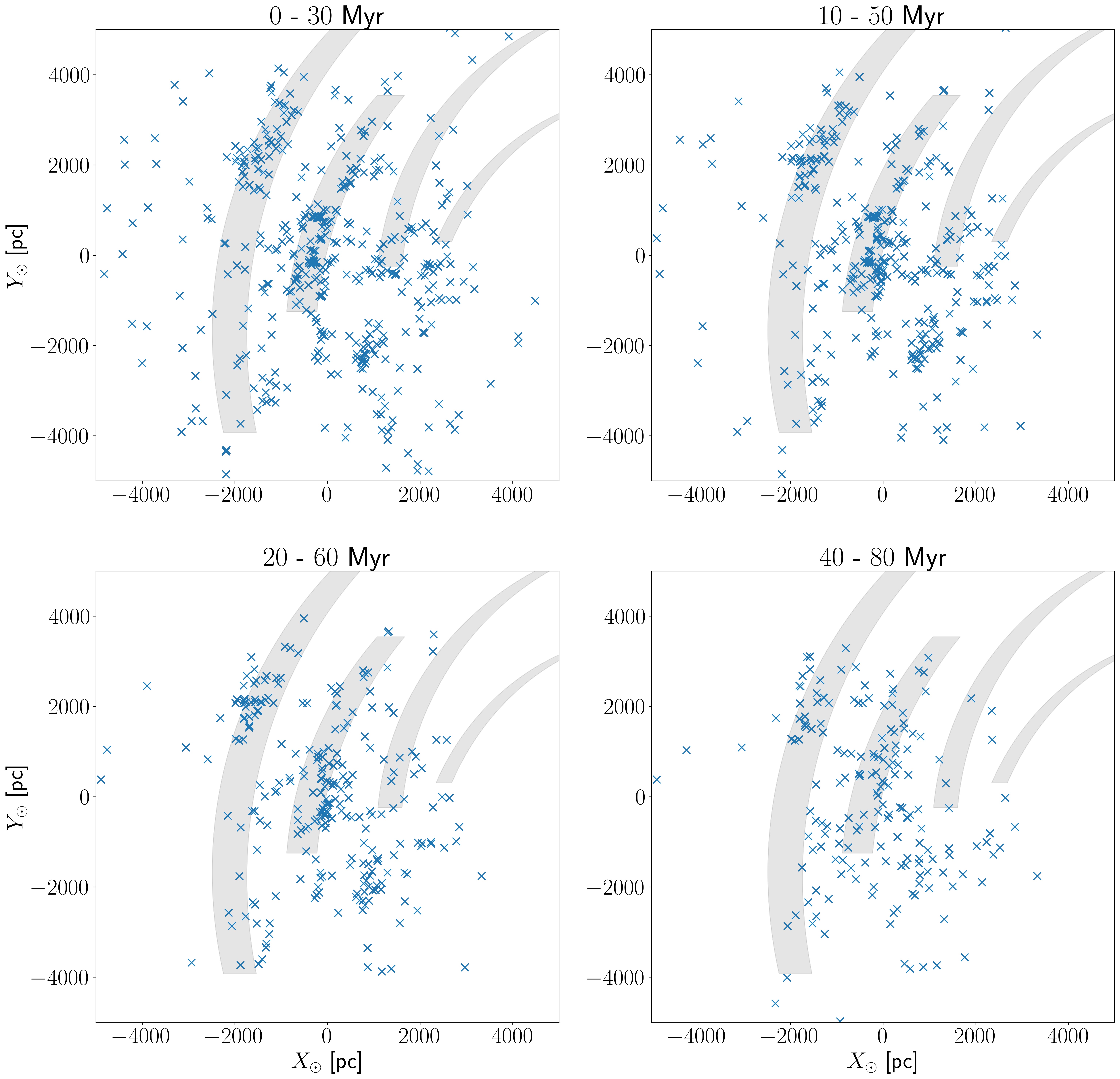}
\caption{Distribution of OCs in heliocentric $X$ - $Y$, in different age bins. The Galactic centre is towards positive $X$ values, and the direction of the Galactic rotation is towards positive $Y$ values. Blue crosses are OC with age estimations, regardless if they have available radial velocity measurements or not, they correspond from left to right and from top to bottom to: ages less than $30$ Myr, from $10$ to $50$ Myr, from $20$ to $60$ Myr and from $40$ to $80$ Myr. The spiral arms defined by \citet{2014ApJ...783..130R} are overplotted.}
\label{fig:xy_distr_age}
\end{figure}

The spiral pattern is clearly seen in a heliocentric $X$-$Y$ projection of the OCs with ages younger than $150$ Myr \citep[see Fig.~8 and Fig.~1 of][respectively]{2020A&A...640A...1C,2020arXiv200407261K} while for older age bins this pattern disappears. Further dividing this $0$-$150$ Myr range in four bins, we are able to spot the OCs overdensities corresponding to the spiral arms in a range of ages. This traces the evolution of the spiral pattern during the time interval where the overdensities are seen (see Sect.~\ref{subsec:estimation_omegap}). \mbox{Figure~\ref{fig:xy_distr_age}} shows how spiral arm segments in the Solar neighbourhood are clearly seen in OC overdensities for the youngest age interval explored ($0$-$30$ Myr), and how these overdensities show an increasing dispersion with time, so a slow dilution. The black shaded regions represent the spiral arms as modeled by \citet{2014ApJ...783..130R}.

\subsection{Re-determination of current spiral arms}
\label{subsec:present_age_arms}

We re-determine the parameters of the present-day spiral arms by using the hypothesis that OCs are born in spiral arms \citep{1969ApJ...158..123R}, and that the youngest OCs ($\leq 30$Myr) have not moved far from their birth places \citep{2005ApJ...629..825D}. Thus, considering the usual log-periodic spiral arms, each arm should be detected as an overdensity following the relation used by \citet{2014ApJ...783..130R} 
\begin{equation}
\label{eq:spiral}
\ln{\frac{R_G}{R_{G,ref}}} = - (\theta_G- \theta_{G,ref}) \tan{\psi},
\end{equation}
where $R_G$ and $\theta_G$ are Galactocentric radius and azimuth along the arm, respectively. And $R_{G,ref}$, $\theta_{G,ref}$ (taken to be near the median value of $\theta_G$) and $\psi$ are a reference Galactocentric radius and azimuth, and the pitch angle for a given arm. The Galactocentric azimuth is taken to be $\theta_G = 0$ on the Sun-Galactic centre line, and growing towards the Galactic rotation direction.

We detect the overdensities using a Gaussian mixture model (GMM) in the $(\ln{R_G},\theta_G)$ space. A GMM is able to describe all the points in the parameter space as a weighted sum of Gaussians. This representation of our sample allows us to describe each arm, expected to follow Eq.~\ref{eq:spiral} with some dispersion, as a Gaussian along that direction (straight line in the $(\ln{R_G},\theta_G)$ space). The number of Gaussians to fit is automatically selected using the Bayesian information criterion (BIC).

\begin{figure}[!htb]
\includegraphics[width = 1.\columnwidth]{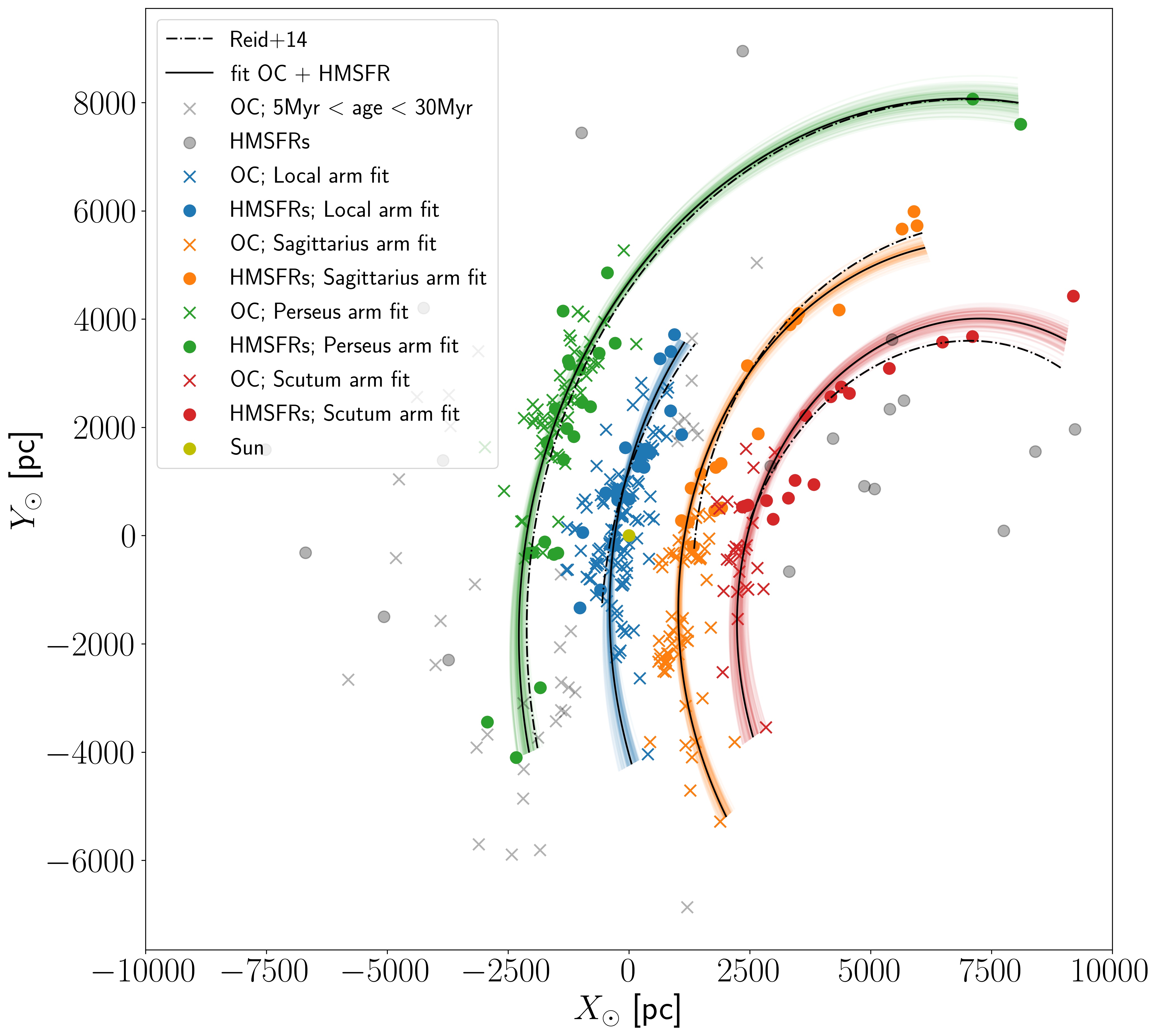}
\caption{Heliocentric $X$ - $Y$ distribution of OCs (crosses) younger than $30$ Myr and HMSFRs (dots) from \citet{2014ApJ...783..130R}, used to fit the spiral arms. Different colours correspond to different arms. The assignments to each arm is computed using a Gaussian Mixture Model. Solid black lines are the fitted spiral arms with the parameters in Table~\ref{tab:spirals}, while shaded regions account for $1\sigma$ uncertainties. Dash-dotted lines correspond to the spiral arms defined by HMSFRs only. The Galactic centre is towards positive $X$ and the Galactic rotation direction is towards positive $Y$.}
\label{fig:today_spiral}
\end{figure}

Once the Gaussian field is obtained, we select the Gaussian components with the four highest weights, corresponding to the four arm segments. We find $56$, $121$, $61$ and $26$ OCs younger than $30$ Myr assigned to the Perseus, Local, Sagittarius and Scutum arms, respectively. To increase the number of spiral arm tracers, and be able to trace the arm in a wider range of Galactic azimuth, we include the data from \citet{2014ApJ...783..130R} used to fit the spiral arms. These data correspond to $103$ high-mass star forming regions (HMSFRs) with parallax and proper motion measurements obtained using Very Long Base Interferometry (VLBI) techniques, $84$ of which are assigned to one of the four explored spiral arms. In order to obtain the parameters for each arm, we fit Eq.~\ref{eq:spiral} to the OCs and HMSFRs assigned to each arm by the minimum least squares method ($348$ tracers in total, $264$ OCs and $84$ HMSFRs). The parameters obtained for each arm are listed in Table~\ref{tab:spirals}.

Figure~\ref{fig:today_spiral} shows the representation of the spiral arms defined by the OCs and HMSFRs. The black solid lines correspond to the best fit value for each arm, and the black shaded regions correspond to the $1\sigma$ uncertainty taken into account the correlations among the estimated parameters. Our all-sky OC sample provides a good complement to the ground-based observations used by \citet{2014ApJ...783..130R}, who do not cover the fourth quadrant. By increasing the number of total tracers by factor of $4$ we can better constrain the estimation of the mean Galactocentric radius $(R_G)$ and pitch angle $(\psi)$ finding lower uncertainties in these values, as well as increasing the $\theta_G$ range where the arms are defined. The spiral arms defined by \citet{2014ApJ...783..130R} using HMSFRs are shown in dash-dotted lines to compare with our definition using both young OCs and HMSFRs.

\begin{table}
\caption{Fitted parameters, including statistical errors, for present-day spiral arms.}
\label{tab:spirals}
\centering
\setlength\tabcolsep{3pt}
\resizebox{\columnwidth}{!}{\begin{tabular}{cccccc}
\hline
\hline 
\multicolumn{1}{c}{Arm} & 
\multicolumn{1}{c}{\begin{tabular}[c]{@{}c@{}}$N_{\text{tracers}}$\\$\text{{\tiny OC+HMSFR}}$\end{tabular}} &
\multicolumn{1}{c}{\begin{tabular}[c]{@{}c@{}}$\theta_{G,\text{ref}}$\\$\text{[deg]}$\end{tabular}} &
\multicolumn{1}{c}{\begin{tabular}[c]{@{}c@{}}$\theta_{G,\text{range}}$\\$\text{[deg]}$\end{tabular}} &
\multicolumn{1}{c}{\begin{tabular}[c]{@{}c@{}}$R_{G,\text{ref}}$\\$\text{[kpc]}$\end{tabular}} &
\multicolumn{1}{c}{\begin{tabular}[c]{@{}c@{}}$\psi$\\$\text{[deg]}$\end{tabular}} \\
\hline
Perseus & $56+24$ & $-13.0$ & $(-20.9,88.2)$ & $10.88 \pm 0.04$ & $9.8 \pm 0.9$ \\
Local & $121+25$ & $-2.3$ & $(-26.9,26.6)$ & $8.69 \pm 0.01$ & $8.9 \pm 1.3$\\
Sagittarius & $61+18$ & $3.5$ & $(-39.3,67.7)$ & $7.10 \pm 0.01$ & $10.6 \pm 0.8$\\
Scutum & $26+17$ & $-4.8$ & $(-32.7,100.9)$ & $6.02 \pm 0.02$ & $14.9 \pm 1.6$\\
\hline
\end{tabular}}
\end{table}

\section{Spiral Pattern Speed}
\label{sec:spiral_patspd}

The spiral pattern speed is indicative of the nature of the spiral arms. As described in \citet{2011MSAIS..18..185G}, the most direct way to estimate the pattern speed of the spiral arms is through the OC population due to the robustness with which their parameters can be estimated, by averaging over their members. With the assumption that the OCs are born in spiral arms \citep{1969ApJ...158..123R}, and integrating backwards the present OCs position to their birthplaces, it is possible to compute the rotation rate at which a spiral arm has moved to reach its present-day position.

To allow for the orbit integration, we use the OC sample with radial velocity measurements available (see Fig.~\ref{fig:OC_sample} for reference). We compute the birthplace of the OCs by integrating backwards in time following each OC orbit. The orbits are integrated following a gravitational potential composed by a spherical nucleus and bulge, a Navarro-Frenk-White dark matter halo and a Miyamoto-Nagai disc, where its parameters have been adapted to follow the observed rotation curve of the Milky Way \citep{2015ApJS..216...29B}. The numerical processing is done using the Python package \texttt{GALPY} \citep{2015ApJS..216...29B}, which uses a Leapfrog integration scheme to trace back the orbits in time steps of $0.1$ Myr. The determination of the uncertainties on the birth position is done via Monte Carlo sampling from the uncertainties on the age of each OC, which is the biggest source of error in our case (see description of the OC sample in Sect.~\ref{sec:data}). The assumed values to normalise the rotation curve for the integration are $R_{\odot} = 8.178$ kpc \citep{2019A&A...625L..10G} and $Z_{\odot} = 20.8$ pc \citep{2019MNRAS.482.1417B} for the Solar position, and the Solar motion of $U_{\odot} = 11.1$, $\Theta_0 + V_{\odot} = 248.5$, $W_{\odot} = 7.25$ km s$^{-1}$ \citep{2010MNRAS.403.1829S,2020ApJ...892...39R}.

It is important to note that the birth position of each OC reveals the location of a spiral arm at a time equals to the birth time of the OC. Similar to the method described in \citet{2005ApJ...629..825D}, the arm at this previous epoch is rotated forward with a rotational velocity equal to the pattern speed of that arm, $\Omega_{p}$, during a time equal to the age of the OC. We consider each arm to have a unique pattern speed, which is constant during the whole time interval considered, and free to be different from other arms pattern speed.

The procedure to compute the pattern speed $\Omega_{p}$ that best describes the data for each arm is as follows: 
\begin{itemize}
  \item Detect overdensities that correspond to the Perseus, Local, Sagittarius and Scutum spiral arms (see Sect.~\ref{subsec:present_age_arms}). Study each arm separately.
  \item Integrate backwards each OC orbit to find their birthplaces. These OC birthplaces represent the location of the spiral arm at the time the OC was born.
  \item Rotate past location of the arm with a circular motion at a given pattern speed ($\Omega_p$) during the age of the cluster ($t$) to find its expected present-day location, \text{i.e.} \mbox{$\theta_{G,\mathrm{now}} = \theta_{G,\mathrm{birth}} + \Omega_p * t$}. 
  \item Iterate over $\Omega_p$ to find the optimal value by minimising the distance of the recovered present-day locations of the spiral arms to their analytical present-day description. The minimisation is done in the present-day to avoid having multiple spiral arm representations for multiple ages (one for each OC).
  \item Repeat procedure for $1\,000$ Monte Carlo realisations to account for the uncertainties in the birthplace of the OC.
  \item For each arm, report the best value for $\Omega_p$ as the mean value of all the pattern speeds obtained, with the standard deviation as its dispersion.
\end{itemize}

\subsection{Test simulation}
\label{subsec:test_simu}

To test our ability to recover the pattern speed, we set up a basic simulation following the evolution of both a density wave spiral pattern and the objects born in it. We generate a log-periodic spiral arm with the parameters taken from \citet[Table~2, Local arm]{2014ApJ...783..130R}, which we take to be the present-day position. In this case, the actual spiral arms are described using the parameters found by \citet{2014ApJ...783..130R}, and not our own estimation (Table~\ref{tab:spirals}); this is because we want to keep the position of the spiral arm and its velocity to be defined by independent tracers (by HMSFR and OCs, respectively). We rotate backwards the arm keeping its shape parameters unchanged, at a constant pattern speed which we assume. At times $T = 10, 20, 30 ,40, 50$ and $60$ Myr, a set of simulated clusters (equivalent to OCs) is generated, which are used as tracers of the spiral pattern speed.

\begin{figure}[!htb]
\begin{subfigure}{1.00\columnwidth}
\includegraphics[width = 1.\columnwidth]{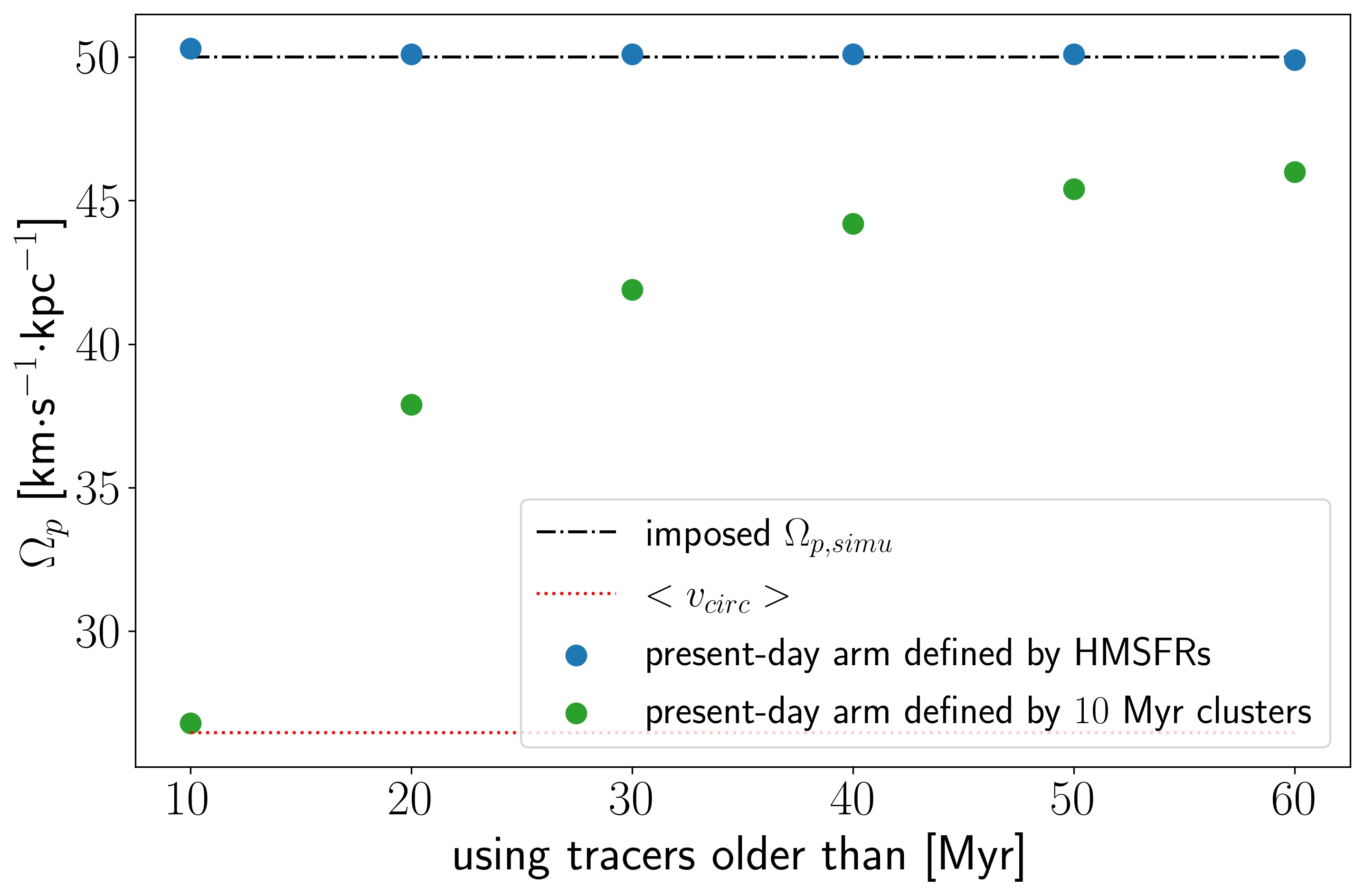}
\caption{Test 1. Imposed $\Omega_p = 50$ $\text{km}\,\text{s}^{-1}\text{kpc}^{-1}$. Blue dots show the recovered $\Omega_p$ value when considering the present-day spiral arm defined by the HMSFRs, while green dots represent the recovered value when the present-day arm is defined with $10$ Myr clusters.}
\label{fig:testa}
\end{subfigure}
\begin{subfigure}{1.00\columnwidth}
\includegraphics[width = 1.\columnwidth]{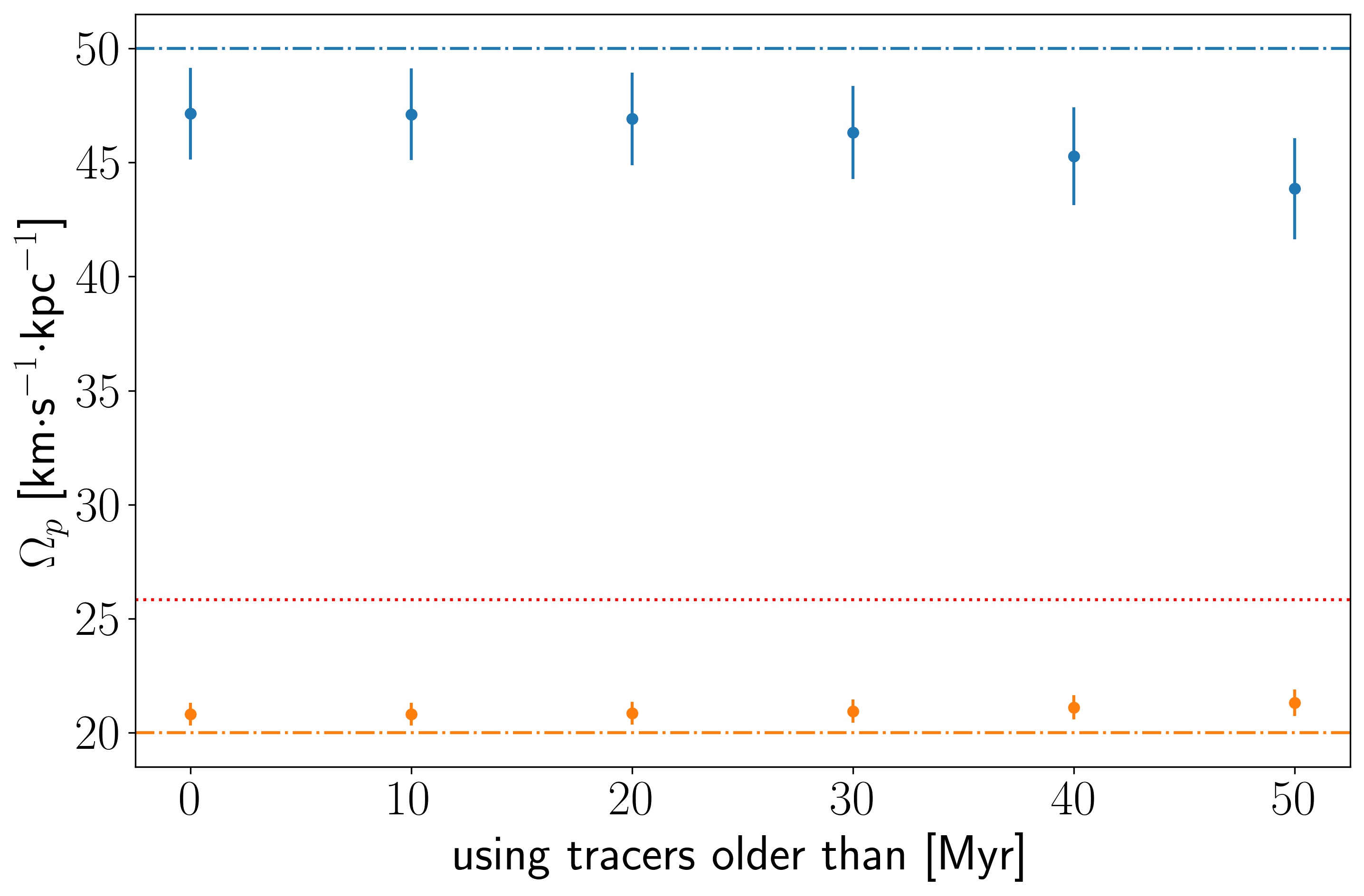}
\caption{Test 2. Pattern speed obtained for the cases of $\Omega_p = 20$ and $50$ $\text{km}\,\text{s}^{-1}\text{kpc}^{-1}$, orange and blue, respectively. Dots show the recovered value, including errorbars.}
\label{fig:testb}
\end{subfigure}
\caption{Recovered $\Omega_p$ for two tests. The $y$-axis represents the value of the pattern speed, and the $x$-axis represents the minimum age of the stellar objects considered for the computation, \textit{i.e.} for $x = 10$ we consider objects with age $\geq 10$ Myr. The dash-dotted lines represent the imposed value and dots represent the recovered value. Dotted red line shows the mean circular velocity of the stellar objects.}
\label{fig:simulation_test_2}
\end{figure}

Firstly, we test the effect of the definition of the present-day spiral arm on the determination of its pattern speed, in the ideal case of the OCs moving with circular orbits (Test~$1$). We let the simulated clusters with different ages evolve to their present-day position with circular orbits using the Milky Way rotation curve from \citet{2015ApJS..216...29B}. In this case, the spiral pattern speed is fixed at $50$ $\text{km}\,\text{s}^{-1}\text{kpc}^{-1}$, while the mean circular velocity of the simulated clusters is $26.37 \pm 1.91$ $\text{km}\,\text{s}^{-1}\text{kpc}^{-1}$. Figure~\ref{fig:testa} shows the capabilities of the method to compute the spiral pattern speed when the present-day spiral arm is defined by i) independent means (HMSFR), or ii) the youngest simulated clusters. In the first case (blue dots), the imposed value for the spiral pattern speed (black dash-dotted line) is always recovered. In the second case (green dots), we recover the value for the circular velocity (red dotted line) when the exact same tracers are used to define the present-day spiral arm and its pattern speed, but we can asymptotically approach the true value when older tracers are considered for the spiral pattern speed computation. From Test~1, we learn that the tracers to define the present-day spiral arms should be independent of the tracers used to compute their pattern speed. This is achieved by defining the present-day spiral arms using the HMSFRs reported in \citet{2014ApJ...783..130R}, which are younger than $10$ Myr, and using OCs older than $10$ Myr to compute the spiral pattern speed.

Secondly, we test if the methodology is able to distinguish two different spiral pattern speeds in a more realistic situation (Test~$2$). The simulated clusters are now evolved using circular velocities and non-zero peculiar velocities, which are drawn from a Gaussian distribution $\mathcal{N}(0,5)$ $\text{km}\,\text{s}^{-1}$. We also add errors to the age of the simulated clusters, consistent with those in our catalogues (see. Sect~\ref{subsec:ages}). The method is run for spiral pattern speeds of $20$ and $50$ $\text{km}\,\text{s}^{-1}\text{kpc}^{-1}$, and assuming that the present-day spiral arm is known from independent tracers, \textit{i.e.} HMSFRs. In Fig.~\ref{fig:testb} we show the obtained values for the two different experiments. We can recover the imposed pattern speed with a systematic error that ranges from $0.8$ to $6$ $\text{km}\,\text{s}^{-1}\text{kpc}^{-1}$ for these different cases of $\Omega_p$. From Test~2, we see that even though the accuracy is not enough to recover the exact individual pattern speeds, the methodology is accurate enough to differentiate between the two scenarios, the $20$ and the $50$ $\text{km}\,\text{s}^{-1}\text{kpc}^{-1}$ cases.

\subsection{Estimation of $\Omega_p$}
\label{subsec:estimation_omegap}

\begin{table*}
\caption{Pattern speed (in $\text{km}\,\text{s}^{-1}\text{kpc}^{-1}$) obtained for different age bins for each spiral arm analysed.}
\label{tab:spiral_pattern_speed}
\centering
\begin{tabular}{ccccc}
\hline
\hline 
Arm & $\Omega_p$ ($10$ - $50$ Myr)& $\Omega_p$ ($20$ - $60$ Myr) & $\Omega_p$ ($40$ - $80$ Myr)  & $\Omega_p$ ($10$ - $80$ Myr)\\
\hline
Perseus & $17.82\pm2.98$ & $18.52\pm2.60$ & $21.66\pm2.57$ & $21.25\pm2.16$ \\
Local & $33.76\pm0.94$ & $34.31\pm1.00$ & $30.92\pm0.85$ & $31.85\pm0.77$\\
Sagittarius & $26.10\pm4.05$ & $29.68\pm3.49$ & $26.47\pm2.18$ & $26.52\pm2.03$\\
Scutum & $49.81\pm1.90$ & $47.67\pm2.39$ & $46.43\pm2.24$ & $46.85\pm1.73$\\
\hline
\end{tabular}
\end{table*}

As seen in the test simulations, the success of the methodology relies on the ability to define the present-day spiral structure by tracers other than OCs, which are used to trace the spiral pattern speed. As already mentioned, we consider that this is achieved by describing the present-day spiral structure with the HMSFRs reported in \citet{2014ApJ...783..130R}, and therefore we use this definition of the present-day spiral arms to compute $\Omega_p$. The authors provide the parameters from a fitting using $84$ HMSFRs, younger than $10$ Myr, with parallax and proper motion measurements from VLBI, as said in Sect.~\ref{subsec:present_age_arms}. This information is available for the Perseus, Local, Sagittarius and Scutum arms. \citet{2019ApJ...885..131R} updated the spiral arms parameters by allowing kinks in the arm which change the pitch angle at a given $\theta_G$. This effect is small for most of the arms we analyse, \textit{i.e.} there is no pitch angle variation in the Local arm and the variations in the Perseus and Scutum arms are within the uncertainties. Since these updates do not lead to significant changes in our calculations, we use the \citet{2014ApJ...783..130R} model for simplicity.

Once the present-day spiral arms are defined, we have to find the present position of the OCs born in each of these arms. In the density wave theory, the OCs may have evolved differently from their mother spiral arms, \textit{i.e} OCs move at a velocity approximately given by the Milky Way rotation curve while the spiral pattern moves at $\Omega_p$. Even though the evolution follows different paths, an overdensity of very young OCs in $(R_G,\theta_G)$ will come from the same arm (see \mbox{Fig.~\ref{fig:xy_distr_age}}). As described in Sect.~\ref{subsec:present_age_arms}, the OCs belonging to each of the arms are selected using a GMM in the $(\ln{R_G},\theta_G)$ space.

We apply the described methodology to the OCs younger than $80$ Myr, for different age ranges to account for the effects seen in Test~1 of Sect.~\ref{subsec:test_simu}. Table~\ref{tab:spiral_pattern_speed} shows the computed spiral pattern speed for the four spiral arms explored, and they show a similar trend as in the test simulation scenario. Following the same argument, we can say that the methodology is good enough to distinguish among different true spiral pattern speeds.

\begin{figure}[!htb]
\includegraphics[width = 1.\columnwidth]{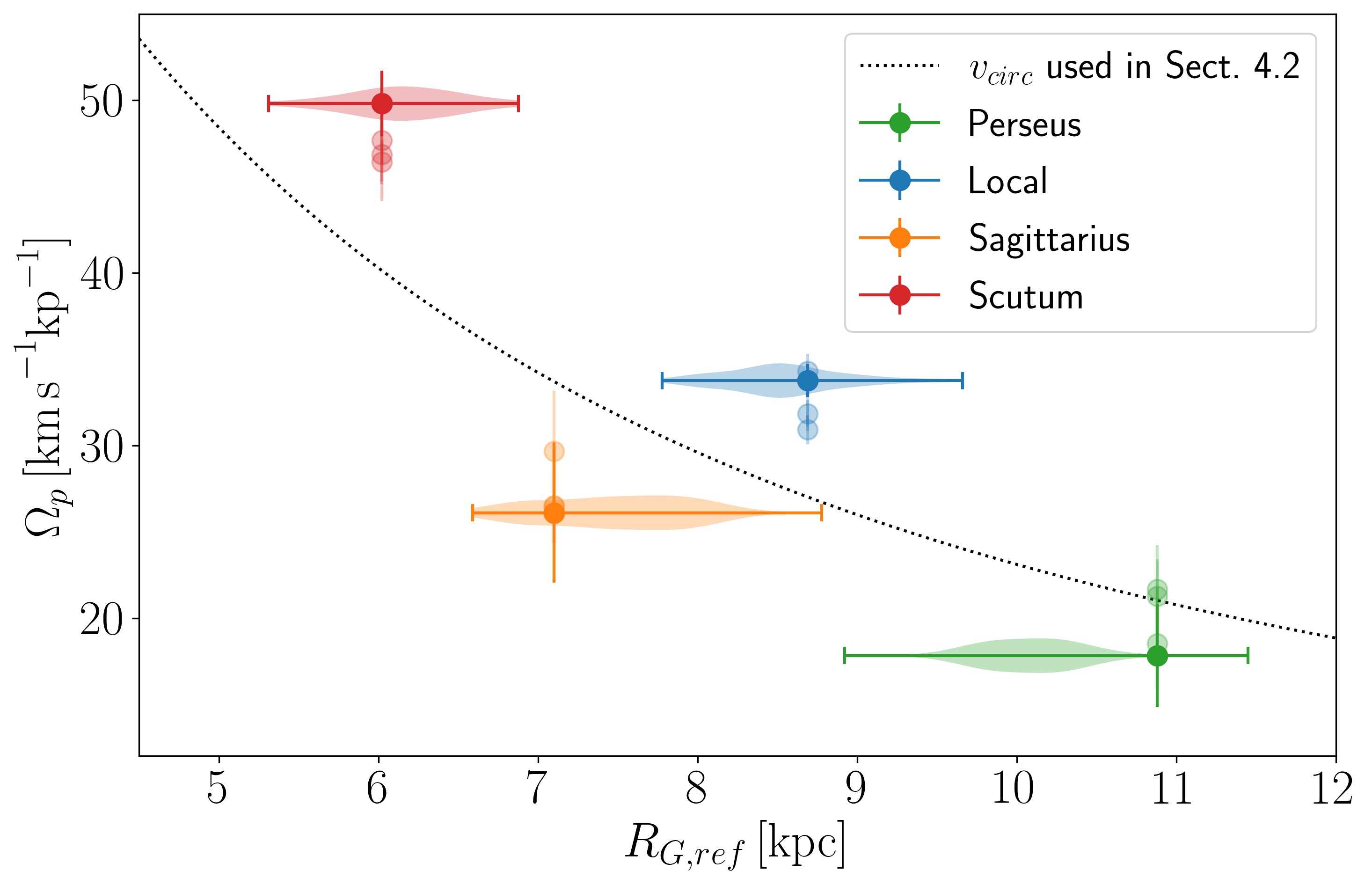}
\caption{Computed spiral pattern speed for the Scutum, Sagittarius, Local and Perseus arms (from left to right). Solid dots show the first column in Table~\ref{tab:spiral_pattern_speed}, corresponding to $10$ - $50$ Myr interval. The transparent dots are for the rest of the columns. The shaded violin plots in the x-axis show the distribution of $R_G$ for the OCs in the $10$ - $50$ Myr interval. The dotted line shows the circular velocity from the Milky Way rotation curve. The estimated $\Omega_p$ values show a decreasing trend with Galactocentric radius.}
\label{fig:pat_spd_radius}
\end{figure}

The recovered values for $\Omega_p$, shown in Fig.~\ref{fig:pat_spd_radius}, are decreasing as the Galactocentric reference radius ($R_{G,ref}$) of the spiral arm increases. The $\Omega_p$ for the explored arms decreases closely following the Galactic rotation curve which is represented by the dotted line, except for the case of the Local arm that breaks this trend. This can be related to the fact that the Local arm could be considered not a long arm but a small armlet or a growing arm instead, however this deserves further study \citep{2016SciA....2E0878X,2017ApJ...835L..18L,2020ApJ...900..186E}. Our results are in agreement with the findings of \citet{2018MNRAS.480.3132Q}, who estimated the spiral pattern speeds for different spiral features to explain the arcs and ridges seen in the velocity distribution of the Solar neighbourhood in {\em Gaia} DR2. The authors studied how the orbits of known moving groups could be perturbed by the presence of a spiral arm, and found that a spiral arm segment in the outer disc located at $\sim 2$ kpc from the Sun, with a pattern speed of $20 \pm 3$ $\text{km}\,\text{s}^{-1}\text{kpc}^{-1}$ could be responsible for the outer boundary of the Sirius/UMa moving group. This finding is in agreement with the spiral speed of the Perseus arm segment we computed using OCs as the main tracers, with a $\Omega_p = 17.82 \pm 2.98$ $\text{km}\,\text{s}^{-1}\text{kpc}^{-1}$ at a Galactocentric radius of $10.88$ kpc. Our results for the rest of the spiral arm segments explored are in a similar agreement \citep[see Table~1 from ][]{2018MNRAS.480.3132Q}, also for the case of the Local arm where the authors found a pattern speed higher than the angular velocity from the rotation curve. The decreasing spiral pattern speed with Galactocentric radius, with spiral arms nearly co-rotating with Galactic rotation, as expected if the spiral arms are short-lived transient structures \citep{2012MNRAS.421.1529G,2014MNRAS.443.2757K}. 

\section{Cluster ages across the spiral arms}
\label{subsec:age_distr}

The analysis of the distribution of the OCs as a function of age across a given arm can provide clues on the nature of the spiral arms \citep{2014PASA...31...35D}, and therefore it offers an independent approach to support the findings of the previous section. As studied by \citet{2010MNRAS.409..396D}, the differences in the rotational velocity of the stellar distribution and the spiral arms, lead to different distributions of the OCs across the present-day spiral arms. Such distributions depend on the spiral arm formation mechanisms. The authors considered a set of four simulations where the spiral structure has been excited by different possible mechanisms, i) a global density wave, ii) a central rotating bar, iii) flocculent spiral or iv) tidally induced arms; and discussed how would be the age distribution of the clusters across a given spiral arm in each of the explored cases. A density wave and/or bar induced spiral arms yield a trend in age across the arms. Flocculent or tidally induced mechanisms yield several individual peaks across the arm, with no age-gradient. This age gradient is due to the difference in velocity with which the different structures (spiral pattern and clusters) are moving, while in the density wave or bar induced spirals scenario the spiral pattern moves with a fixed, constant pattern speed, the clusters move following the galactic rotation curve. That results in older (younger) clusters leading the spiral arm, if the clusters are inside (outside) the co-rotation radius. In the opposite case, for the flocculent and tidally induced arms where the spiral pattern and the stars move at roughly the same speed, the section of the arm contains clusters of different ages with no clear gradient across the arm.

\begin{figure*}[!htb]
\includegraphics[width = 1.\textwidth]{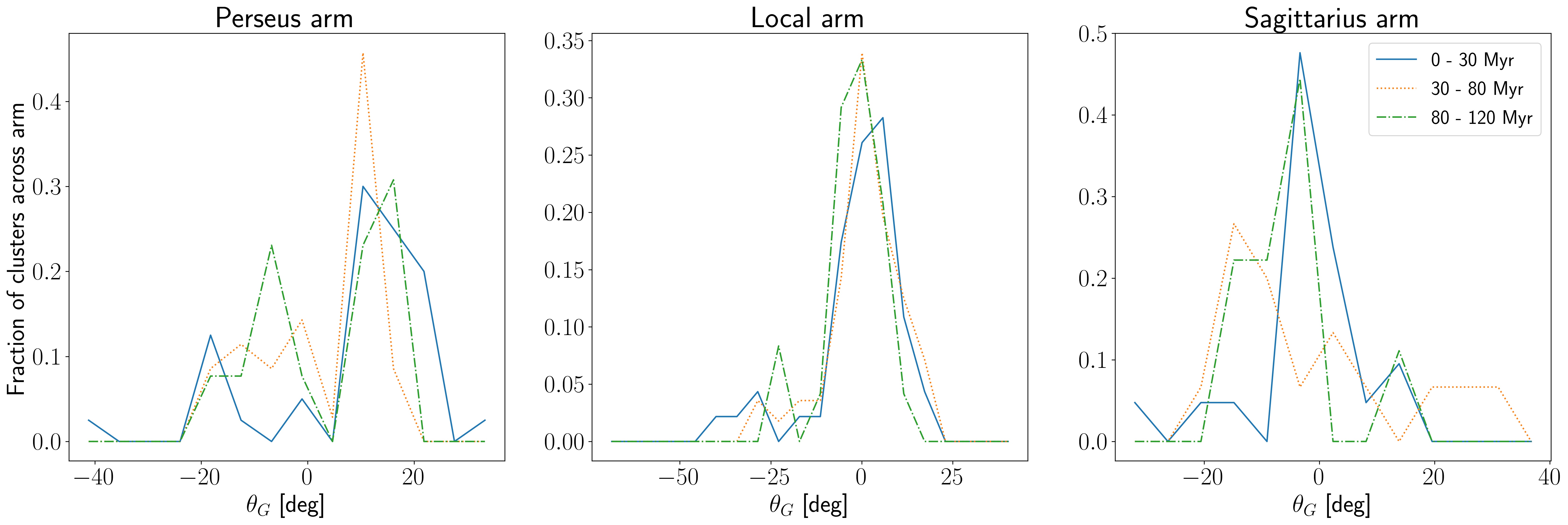}
\caption{Fraction of OCs for different age bins across the Perseus (left), Local (middle) and Sagittarius (right) arms. The $x$-axis represents the Galactic azimuth ($\theta_G$) explored for a circular section centred at the Galactic centre, at a distance of $10$ (left), $8.3$ (middle) and $7$ (right) kpc. The $y$-axis gives the number of OCs in each $\theta_G$ bin over the total number of OCs in the whole circular section. Solid blue lines, dotted orange lines and dash-dotted green lines correspond to clusters in $0$-$30$ Myr, $30$-$80$ Myr and $80$-$120$ Myr age bins. The number of OCs in each panel for the $0$-$30$ Myr, $30$-$80$ Myr and $80$-$120$ Myr age bins is: 40, 35 and 13 for the Perseus arm; 46, 56 and 24 for the Local arm; and 21, 15 and 9 for the Sagittarius arm, respectively.}
\label{fig:dobbs_1}
\end{figure*}

In Fig.~\ref{fig:dobbs_1}, we show a plot reproducing Fig.~4 of \citet{2010MNRAS.409..396D} but using our OCs sample. The different panels show histograms of the number of OCs, for different cluster ages, across a circular section ($500$ pc wide) located at distances of $10$ kpc, $8.3$ kpc and $7$ kpc from the Galactic centre, \textit{i.e.} approximately along the Perseus, Local and Sagittarius arm, respectively. None of the arms shows the aforementioned age gradient, which should be clearer as we move away from a hypothetical co-rotation radius. This indicates that the velocity of the clusters, \textit{i.e.} the stellar distribution velocity, is very similar that the rotation velocity of the spiral arms, therefore co-rotating with them. We have explored different sizes of the age bins reaching the same conclusion in all cases. The non-presence of the age gradient favours the flocculent spirals or the external tidal interaction, where spiral arms tend to be transient, as the mechanisms for the excitation of the spiral structure.

The completeness of the OC population may play a role in the interpretation of Fig.~\ref{fig:dobbs_1}. \citet{2020A&A...635A..45C} tested how many known (prior to {\it Gaia}) OCs could be recovered using their detection algorithm. This recovery fraction was then used by \citet{2020arXiv200601690A} to estimate the completeness of the OC population as a function of age, finding that the recovered fraction of OCs is $\lesssim 60\%$ for the very young OCs (in the range of $1$-$10$ Myr), and reaching $\gtrsim 90\%$ for older OCs. Therefore, even if the fraction of youngest population in Fig.~\ref{fig:dobbs_1} may be underestimated, the older populations (defining the age gradient) are nearly complete, reaching the same conclusions of no age gradient at all.

\section{Discussion}
\label{sec:discussion}

The nature of the spiral arms has been studied in other galaxies taking advantage of our external point of view. \citet{2018MNRAS.478.3590S} studied the distribution of stellar clusters across the spiral arms in NGC~$1566$, M~$51$ and NGC~$628$. They find an age gradient across the arm only in NGC~$1566$ (a grand design spiral galaxy with a strong bar), which is compatible with the density wave scenario \citep{2010MNRAS.409..396D}. For the case of M~$51$, the spiral structure is excited by the tidal interactions with its companion, and for NGC~$628$ the spiral arms are consistent with a pattern speed decreasing with radius, both leading to a transient spiral nature. Also in external galaxies, the measurements of spiral pattern speeds that vary as a function of radius \citep{2008ApJ...688..224M,2012ApJ...752...52S} or the evolution of the spiral arms pitch angle \citep{2019MNRAS.490.1470P}, support a transient nature for their spiral structure. This transient behaviour of the spiral structure, with the arms co-rotating with disc stars, is also expected from N-body simulations for unbarred galaxies or galaxies with a weak bar \citep{2013MNRAS.432.2878R}; while galaxies with a strong bar quickly develop a spiral pattern whose pattern speed is constant with radius, behaving as a global density wave as for the case of NGC~$1566$.


For the case of the Milky Way, the lack of a homogeneous OC catalogue (before {\em Gaia}) prevented from reaching a firm conclusion \citep{2017MNRAS.466.3636M}. Thanks to the {\em Gaia} mission, the study of OCs has reached a maturity, in terms of purity and homogeneity of the catalogue, and robustness of its estimated parameters, that allows us to apply different approaches to revisit the spiral nature of the Milky Way.

We have explored the nature of the spiral structure of the Milky Way by comparing the angular velocities in which the stellar distribution and the spiral pattern move. The spiral arms should move with a global constant pattern speed in the density wave scenario, regardless of the Galactocentric reference radius of the arm. This is not what we deduce from our sample of young OCs as main tracers. We see that different spiral arm segments move with a different angular velocity, which tend to decrease with their Galactocentric reference radius. This behaviour is related to a short-lived transient spiral structure. 

The procedure applied in this work to compute the spiral pattern speed uses the hypothesis of spiral arms with a constant shape during the time interval explored. Our tests using simulations show that the methodology is accurate enough to discard a unique pattern speed for all the spiral arms studied. The computed $\Omega_p$ values for the arms are discrepant in more than $5\sigma$ for all the arms except for the case of the Sagittarius with Local arms and Sagittarius with Perseus arms, where the discrepancy is of $1.8\sigma$ and $1.6\sigma$, respectively. For these cases, the uncertainty weighted mean of the $\Omega_p$ values is $32.03 \pm 0.87$ $\text{km}\,\text{s}^{-1}\text{kpc}^{-1}$. Therefore, we consider each arm to have its own spiral pattern speed, which is also supported by the age gradient of the OC seen across the spiral arms studied in Sect.~\ref{subsec:age_distr}. This is in contrast with the work done by \citet{2019MNRAS.486.5726D} who, using the same methodology, reported a spiral pattern speed of $\Omega_p = 28.2\pm2.1$ $\text{km}\,\text{s}^{-1}\text{kpc}^{-1}$, common for all the explored spiral arms. Here, the inclusion of hundreds of newly discovered OCs \citep{2020A&A...635A..45C}, with an updated estimation of ages, distances and line-of-sight extinctions for the whole OC sample \citep{2020A&A...640A...1C} and the addition of radial velocities for a large fraction of them \citep{2020arXiv201204017T}, together with a robust statistical treatment, allows us to distinguish among different true pattern speeds for different spiral arms (Fig.~\ref{fig:pat_spd_radius}, Table~\ref{tab:spiral_pattern_speed}). 

The effects that may change the shape of the spiral arms (\textit{e.g.} the shear of the Galactic disc or the evolution of the pitch angle) are not included in the assumptions of this work, nor in the works using similar procedures \citep{2005ApJ...629..825D,2015MNRAS.449.2336J,2019MNRAS.486.5726D}. However, if we consider that these effects are small over the course of $\sim 50$ Myr, the values obtained for the spiral pattern speeds suggest that spiral arms are structures that co-rotate with stars at any radii, revealing a transient nature of these arms \citep{2012MNRAS.421.1529G}. Therefore, our results with OCs agree with other works dealing with the kinematic substructure in the solar neighbourhood. These works, some including simulations, tend to explain the kinematics of moving groups, or features in the action-angle space, with a transient behaviour of the Galactic spiral arms \citep{2018MNRAS.480.3132Q,2020arXiv200601723Q,2018MNRAS.481.3794H,2019MNRAS.484.3154S}.

In addition, we have explored the imprint in the age-distribution of the OCs across the spiral arms, and we do not see the predicted age gradient of density wave or bar-driven spiral arms \citep{2010MNRAS.409..396D}, even when the effects due to the incompleteness of our OC sample are taken into account. The combination of both results allow us to favour a flocculent Milky Way with transient spiral arms, disfavouring the density wave scenario with a grand design morphology. This idea of a flocculent Milky Way was already studied by \citet{2002AJ....124..924Q}, who found multiple spiral features, each with a different pattern speed which is decreasing with Galactocentric radius. From the morphology of the spiral arms, a flocculent Milky Way was favoured by \citet{2016SciA....2E0878X} due to a long Local arm located between the Perseus and Sagittarius arms that would not be explained by a density wave theory with a pure grand design morphology. N-body simulations are also in agreement with density waves not explaining the spiral structure in our Galaxy \citep{2015ApJ...800...53H}.

\section{Conclusions}
\label{sec:conclusions}

We have analysed the OC population with {\it Gaia} EDR3 astrometric parameters, radial velocities compiled from different surveys, and astrophysical parameters computed from {\it Gaia} DR2 astrometry and photometry, in order to derive the structure of the spiral arms in the Solar neighbourhood and to discriminate among several hypothesis about their nature.

We show that each of the four investigated arms exhibits a different pattern speed. Using a combination of statistical and data mining techniques, we find that each spiral arm has a spiral pattern speed which tend to decrease with Galactocentric radius, following the Galactic rotation curve, favouring a transient behaviour for these arms.

We analyse the age-distribution of the OC population across the spiral arms to see the imprint of the different angular velocities of the stellar distribution and the spiral arm segments, if any. We see no indication of the age gradient predicted by \citet{2010MNRAS.409..396D} to be a sign of a density wave-like footprint, thus favouring a flocculent Milky Way.

These two independent experiments, based on the most complete OC sample to date, allow us to disfavour the density wave theory of spiral structure and point towards a transient nature of the spiral arms. This behaviour is seen here for the first time using OCs data, due to the increase in the OC sample with radial velocities available and better estimation of ages and distances. This points towards the same direction as the conclusions obtained by other authors by comparing {\it Gaia} DR2 kinematic information in the solar neighbourhood with simulations including different kinds of spiral arms, representing an agreement on the results using these two different (complementary) datasets. 

Given the transient nature of the spiral arms proposed here, where the stellar objects co-rotate with the arm at any radius, we can increase the number of tracers of these spiral arms by adding the youngest OCs ($\leq 30$ Myr) to the HMSFRs used to define the present-day arms. As a result, we increase by $314\%$ the number of tracers \citep[adding $264$ OCs to the $84$ HMSFRs used in][]{2014ApJ...783..130R}, and report updated parameters for the Perseus, Local, Sagittarius and Scutum spiral arms, spanning a wider range in Galactic azimuth (Table~\ref{tab:spirals}).

\begin{acknowledgements}

The authors thank the referee for his/her constructive comments that helped improve the paper.

ACG thanks Dr. Lennart Lindegren, Dr. Teresa Antoja, Dr. Francesca Figueras, Dr. Maria Mongui\'o and Dr. Pau Ramos for their useful suggestions and comments.
ACG also thanks Dr. Louise Howes for her comments on the writing.

This work has made use of results from the European Space Agency (ESA)
space mission {\it Gaia}, the data from which were processed by the {\it Gaia
Data Processing and Analysis Consortium} (DPAC).  Funding for the DPAC
has been provided by national institutions, in particular the
institutions participating in the {\it Gaia} Multilateral Agreement. The
{\it Gaia} mission website is \url{http: //www.cosmos.esa.int/gaia}. The
authors are current or past members of the ESA {\it Gaia} mission team and
of the {\it Gaia} DPAC.

This work was (partially) supported by the Spanish Ministry of Science, Innovation and University (MICIU/FEDER, UE) through grants RTI2018-095076-B-C21, ESP2016-80079-C2-1-R, and the Institute of Cosmos Sciences University of Barcelona (ICCUB, Unidad de Excelencia ’Mar\'{\i}a de Maeztu’) through grants MDM-2014-0369 and CEX2019-000918-M. 
ACG acknowledges Spanish Ministry FPI fellowship n. BES-2016-078499.
PM gratefully acknowledges support from a research project grant from the Swedish Research Council (Vetenskapr\aa det).
FA is grateful for funding from the European Union's Horizon 2020 research and innovation programme under the Marie Sk\l{}odowska-Curie grant agreement No. 800502 H2020-MSCA-IF-EF-2017.

This work has received funding from the European Union’s Horizon 2020 research and innovation programme under the Marie Skłodowska-Curie grant agreement H2020-MSCA-COFUND-2016-754433.

This work has been supported by the Spanish Government (SEV2015-0493), by the Spanish Ministry of Science and Innovation (contract TIN2015-65316-P), by Generalitat de Catalunya (contract 2014-SGR-1051).

This research has made use of the VizieR catalogue access tool, CDS,
Strasbourg, France. The original description of the VizieR service was
published in A$\&$AS 143, 23.
\end{acknowledgements}

\bibliographystyle{aa} 
\bibliography{bibliography}


\end{document}